\documentstyle[preprint,aps]{revtex}

\begin{document}

\draft
\title{A search for stable strange quark matter nuggets in helium
}

\author{J. Vandegriff,$^{(a)}$ G. Raimann,$^{(a)}$  R.N. Boyd,$^{(a,b)}$
M. Caffee,$^{(c)}$ B. Ruiz$^{(c)}$
}
\address{$^{(a)}$ Department of Physics, The Ohio State University,
Columbus, OH 43210, USA}
\address{$^{(b)}$ Department of Astronomy, The Ohio State University,
Columbus, OH 43210, USA}
\address{$^{(c)}$ Lawrence Livermore National Laboratory,
Livermore, CA 94550, USA}

\maketitle

\begin{abstract}

A search for stable strange quark nuggets has been conducted in helium and
argon using a high sensitivity mass spectrometer. The search was guided by
a mass formula for strange quark nuggets which suggested that stable
strange helium might exist at a mass around 65 u. The chemical similarity
of such ``strangelets'' to noble gas atoms and the gravitational
unboundedness of normal helium result in a large enhancement in the
sensitivity of such a search. An abundance limit of no more than $2 \cdot
10^{-11}$ strangelets per normal nucleus is imposed by our search over a
mass region from 42 to 82 u, with much more stringent limits at most
(non-integer) masses.

\end{abstract}

\newpage

\section{Introduction}

Witten\cite{Wit84} suggested in 1984 that strange matter consisting of up,
down, and strange quarks might be stable. A previous study of such
``strangelets'' by Chin and Kerman\cite{Kerxx} had suggested the
possibility that such quark matter might actually be at least a metastable
phase of nuclear matter. A mass formula was developed in the context of
quarks confined to a (nuclear) bag by Farhi and Jaffe\cite{Far84} and by
Berger and Jaffe\cite{Ber87}; their formula depended only on the bag
constant and the strange quark mass in the nuclear medium. It was used to
study the possible stability of strangelets as a function of baryon number,
as well as possible decay modes. Those studies confirmed the possibility
that strangelets might be stable, although the most likely masses for
stable strangelets were suggested by the mass formula as being very large,
possibly having baryon numbers larger than the heaviest stable nuclei.

However, Takahashi and Boyd\cite{Tak88} extended the Farhi-Jaffe-Berger
mass formula to include shell effects. The magic numbers for quarks in a
strangelet were calculated in the spherical cavity model of DeTar and
Donoghue\cite{Tar83}. Configuration mixing and deformation effects were
included using prescriptions of Myers and Swiatecki\cite{Mye66}. The
resulting corrections were then added to the Farhi-Berger-Jaffe mass
formula. The mass formula indicated that a charge 2 strangelet has a
reasonable chance of being stable, and, if so, would be expected to have a
mass around 65 u. Furthermore, this mass prediction does not depend
critically on the shell correction parameter, although the boundedness of
the resulting strangelets certainly does.

While the possible existence of strangelets is of obvious interest in its
own right, it has also been shown\cite{Boyxx} that they could have
interesting astrophysical consequences. At an abundance of 1 in $10^{15}$
normal nuclei, strangelets could, if they existed with charge 2, affect
stellar evolution. This is due to the greatly reduced Coulomb barrier they
would have for hydrogen burning compared, e.g., to the nuclei involved in
CNO cycle burning. Furthermore, heavy nuclides, such as strangelets, could
have settled to the cores of stars so that a strangelet abundance level
sufficient to affect stellar burning might correspond to a significantly
lower mean cosmic abundance.

Because of the tendency for stable strangelets to be very massive, the
existence of light strangelets in the cosmos may well depend on some sort
of processing mechanism. This was studied by Boyd and Saito\cite{Boyxx}.
They concluded that if massive strangelets are produced as high energy
particles in, e.g., collisions of neutron stars, as was suggested by
Witten\cite{Wit84}, the collisions of the high energy strangelets with the
particles of the nebula which resulted from the supernovae that produced
the neutron stars could well do the processing necessary to spall off many
light strangelets. Furthermore, this possibility appears to be consistent,
both in strangelet mass and energy, with the possible high energy
strangelets seen in the cosmic ray experiments of Saito et al.\cite{Sai78}.

It should be noted that even the possibility for production of strange
objects is controversial. Alcock and Farhi\cite{Alcxy} studied possible
strangelet production in the big bang, but concluded that any remnant
strangelets, unless they were extraordinarily massive, would have
evaporated shortly after the big bang. This, however, has been disputed by
Madsen et al.\cite{Madxy}. Witten\cite{Wit84} suggested that strangelets
could result from collisions of collapsed strange stars. However,
Madsen\cite{Madxx} and others concluded this to be an impossibility, due to
the need for a fairly thick crust on the surfaces of neutron stars so as to
explain pulsar glitches\cite{glitchxx}. However, the structure of strange
stars has been shown by others\cite{Heisxx} to be quite capable of
producing the sort of crust necessary to explain the glitches.

In the present study we have adopted a purely experimentalist approach,
ignoring the theoretical controversies that exist and designing a search
for strangelets that takes advantage of terrestrial selection processes to
enhance their abundance. Searches have been conducted previously (see
\cite{hemmxx} and references therein), most
notably on a range of nuclides including H, Li, Be, B, C, O, F, and Na to
extremely low abundance limits: $10^{-15}$ to $10^{-25}$ strangelets per
normal nucleus. That search, however, did not include He, for which special
properties exist. The present search, guided by the Takahashi-Boyd mass
formula, was based on the hypothesis that stable strange helium nuclei,
i.e., charge 2 strangelets, exist.  The sensitivity of this search relies
on the noble gas chemistry such an entity would have, coupled with the
abundance enhancement it would have as a terrestrial helium ``atom''.

If charge 2 strangelets exist, then the Earth would effectively enrich the
strangelet to He ratio. Neither isotope of helium is gravitationally bound
to the Earth, nor does He interact chemically to form molecules that would
be bound (as does H). However, strangelets possessing the 65 u mass
suggested by the Takahashi-Boyd mass formula would be gravitationally bound
over the time of formation of the Earth's present atmosphere. This effect
greatly enhances the limit obtained by our experimental search.  The
special features involving this type of helium search were described by
Boyd and Caffee\cite{Boyxy}.

Section II describes the noble gas mass spectrometric facility at Lawrence
Livermore National Laboratory where the experiment was performed and also
estimates the enhancement in terrestrial helium that results from the
gravitational unboundedness of ordinary helium. Section III describes the
data analysis and results, and section IV summarizes the results of our
study.

\section{Experimental Details}

\subsection{Mass spectrometric facility and data acquisition}

The spectrometer used in this experiment consists of two major components:
the gas aliquotting and purification system and the mass spectrometer
itself. The spectrometer in turn is comprised of an electron bombardment
ionization source, a magnetic sector, and detectors -- an axial
pulse-counting electron multiplier and an off-axis faraday cup.

Known amounts of noble gases were aliquotted using a combination of
calibrated volumes and capacitive manometers. After a known amount of air
was aliquotted, the noble gases (which are trace constituents) were
purified from air using a titanium sublimation pump and a series of SAES
non-evaporative alloy getters. After purification, the noble gases were
separated from each other using cryogenic techniques.

After purification and separation, the noble gas species of interest was
admitted into the spectrometer, which was run in a static mode. Singly
ionized species were accelerated to $\sim 4.5$ keV and were then mass
analyzed by a standard double focusing magnetic sector. The beam finally
passed through object slits placed directly in front of the axial detector.
The mass resolving power ($M/\Delta M$) of this instrument using the axial
detector is about 600. There is also an off-axis faraday cup for the
measurement of higher ($\gtrsim 10^5$ counts sec$^{-1}$)
beam currents, which measures current via an ultra-stable sensitive
electrometer. The magnetic field was controlled by a Hall probe field
regulator and had a stability of the order of ppm. Operationally there were
$>20$ discrete DAC settings at the peak of each mass. The entire system was
controlled by a Macintosh computer running LabView. A scan of a particular
mass range was accomplished by changing the magnetic field setting of the
magnet in discrete steps and, after each step, acquiring data for a
predetermined period of time. The data were then stored on disk for later
retrieval.

\subsection{Enhancement of measured abundance limits}

The measured strangelet abundance limit in air will be enhanced over the
cosmic value due to the gravitational unboundedness of the primordial
terrestrial helium. Assume that the abundance of noble gas $x$ in air,
$x_a$, is characteristic of its cosmic abundance $x_c$. If $x$-atoms are
gravitationally bound, the ratio $x_a$/$x_c$ would be expected to be
roughly the same for different gases $x.$
For the two heaviest stable noble gases, Xe and Kr,
(with $x_a$ expressed as a volume content fraction \cite{handbook}
and $x_c$ in units of atoms per $10^6$ silicon atoms \cite{andersxx})
\begin{displaymath}
\mbox{Xe}_a/\mbox{Xe}_c  = 1.9 \cdot 10^{-8}
\end{displaymath}
and
\begin{displaymath}
\mbox{Kr}_a/\mbox{Kr}_c = 2.5 \cdot 10^{-8}  \quad ,
\end{displaymath}
confirming our assertion. For Ar, the ratio is
$9.2 \cdot10^{-8}$, for Ne it falls to $5.3 \cdot 10^{-12}$, and for
He to $1.9 \cdot 10^{-15}$. The amount by which this last ratio differs
from that for the heaviest noble gases, $1 \cdot 10^7$, gives the
enhancement for our experiment. It should be noted that most of
terrestrial helium is not that which existed at the time the Earth was
formed, but has been produced since that time by decay of transuranic
nuclides.

Our experiment determined the ratio in air of charge 2 strangelets to
$^4$He. What is desired, however, is the cosmic value of that ratio,
$\mbox{Str}_c/^4\mbox{He}_c$, given by
\begin{displaymath}
\frac{\mbox{Str}_c}{^4\mbox{He}_c}
=\frac{\mbox{Str}_a}{^4\mbox{He}_a}
 \left[ \frac{^4\mbox{He}_a}{^4\mbox{He}_c}
        \frac{\mbox{Xe}_c}{\mbox{Xe}_a} \right]
\end{displaymath}
all values of which are known or determined in our experiment. The inverse
product of the term in brackets gives the $10^7$ enhancement factor (in
addition to a correction for units) due to the gravitational unboundedness
of $^4$He.

This argument depends on the assumption that there are no components in the
mass spectrometer that would select strangelets over $^4$He.  Since there
are numerous cryopumps in the spectrometer, one must consider the effects
on the boiling point of changing the nuclear mass in helium from 4 to 65 u.
This is described in detail by Wilks \cite{wilks}, who gives the procedure
for determining the boiling point for $^{65}$He; we calculated it to be 7
K.  Thus the cryopumps in the system, which operate at a variety of
temperatures, but always above 10 K, should not affect the strangelet to
$^4$He abundance ratio.

The helium ``supply'' we used was just the helium from air, isolated in the
spectrometer by freezing out all other gases.  Thus no additional caveats
over those discussed above are necessary. We also ran a series of
experiments using Argon obtained from air. The Argon was isolated by taking
advantage of its unique adsorption on activated charcoal at liquid nitrogen
temperatures. With only argon adhering to the charcoal, all other gases
could be pumped out. Then with the pump closed off, the charcoal is heated
to expel the argon.  Only argon (and no helium or strange matter) would be
expected to survive this adsorption isolation, because, although
strangelets are closer in mass to argon than helium, the adsorption depends
on electronic properties, such as boiling point, polarizability, etc.

The ratio of heavy to regular helium on Earth should be well represented by
its concentration in air at sea level. The height above the Earth at which
the gas is sampled has no bearing on this ratio, because all atmospheric
gases, even heavy ones such as CO$_2$, are well mixed up to about 80 km
\cite{morgan}. Also, heavy helium would not be expected to settle to the
bottom of the ocean. The oceans recirculate their deep water within
atmospheric contact in less than 1000 years \cite{togxx}, a short time
compared to the age of the Earth, which allows equilibrium to be
established between the helium in the oceans and the atmosphere.  Finally,
the oceans do not represent a significant reservoir of helium, since the
solubility of helium is very low \cite{keesom}. Thus we conclude that a
measurement of the ratio of heavy to regular helium in air should not be
affected by any sinks in the heavy helium.

\section{Data analysis}

Mass spectra were accumulated in successive runs with helium, argon, and
then a background run with no gas.  Data were collected by stepping through
the allowed range of the magnetic field, pausing for a preset time (4 sec)
at each setting to count the number of particles hitting the multiplier.
Using a mass calibration based on known peaks in the spectrum (CO$_2$ at
mass 44 and C$_6$H$_6$ at mass 78), we obtained a counts vs.\ mass spectrum
for each case. While we had hoped to do a direct subtraction of the
background spectra from the helium spectra, this was not possible due to
the somewhat random nature of the background, which caused the peaks to
have slightly different heights and widths in the different spectra.
However, strangelet candidates can be ruled out at some level by visual
inspection; an anomalous peak in the helium spectrum which also appears at
comparable strength in the argon spectrum, several of which were observed,
is not likely to be a strangelet.

Figure \ref{fig1} shows the data and results for the mass scan from 42 to
82 u. The dominant peaks are hydrocarbon contaminants, which appear mostly
at integer masses, but also at half integer masses for doubly stripped
hydrocarbons. Since the counting interval is four seconds and the
efficiency of the multiplier is 75\%, the lowest possible counting rate is
$(1/4)/0.75 = 0.3333$ counts per second, as seen on the graph.

The upper limit for the strangelet abundance detected by our experiment is
derived by assuming that all the events are due to strangelets. To compute
the abundance limit, the count rate at a particular mass value is divided
by the known amount of regular helium in the beam of the spectrometer
($\sim 5 \cdot 10^6$ per second) and the enhancement factor of $1 \cdot
10^7$. Another factor of $11.3$ in the denominator converts the abundance
of strangelets relative to the abundance of cosmic helium (8.9\% number
fraction) to the abundance of strangelets compared to all nuclei
\cite{andersxx}.

For the first and second halves of the mass region, the maximum count rates
are $\sim\!\!10^4$ and $\sim\!\!10^3$, with corresponding abundance limits
for these overall regions of $2 \cdot 10^{-11}$ and $2 \cdot 10^{-12}$
strangelets per nucleus.  Away from the mass peaks at integer and half
integer values, the maximum count rate drops to no more than 10, yielding a
lower limit of $2 \cdot 10^{-14}$ strange to normal nuclei in these mass
regions. This value cannot be quoted for the entire region, however, since
the amount of strangelets present in the contaminant peaks is unknown.

\section{Summary}

If quark matter is present in the universe, then charge 2, helium-like
strangelets might exist at some level in the terrestrial atmosphere.  We
have performed a sensitive mass spectrometric search for strangelets from
42 to 82 u.  No definitive strangelet candidates were observed in our data,
but upper limits on the cosmic strangelet to normal nucleus abundance are
deduced to be $2 \cdot 10^{-11}$ for masses in the 42 to 60 u range, and $2
\cdot 10^{-12}$ for masses between 60 and 82 u.

\acknowledgements

Helpful discussions with J. Tough, D. Edwards, and K. Foland are gratefully
acknowledged.  This work was supported by NSF grant PHY-9221669 and by DOE
grant W-7405-ENG-48.

%%%%%%%%%%%%%%%%%%%%%%%%%%%%%%%%%%%%%%%%%%%%%%%%%%%%%%%%%%%%%%%%%%%%%%%%%%%

%
%
\newpage
%%%%%%%%%%%%%%%%%%%%%%%%%%%%%%%%%%%%%%%%%%%%%%%%%%%%%%%%%%%%%%%%%%%%%%%%%%%

\begin{figure}
\caption{Normalized data, and also final abundance limits as determined by
our experiment.  Peaks at all integer values and some half-integer values
indicate the presence of hydrocarbon contaminants.  For a), a reasonable
upper limit for the abundance is $2 \cdot 10^{-11}$, and for b), the limit
is $2 \cdot 10^{-12}$. }
\label{fig1}
\end{figure}

%%%%%%%%%%%%%%%%%%%%%%%%%%%%%%%%%%%%%%%%%%%%%%%%%%%%%%%%%%%%%%%%%%%%%%%%%%%

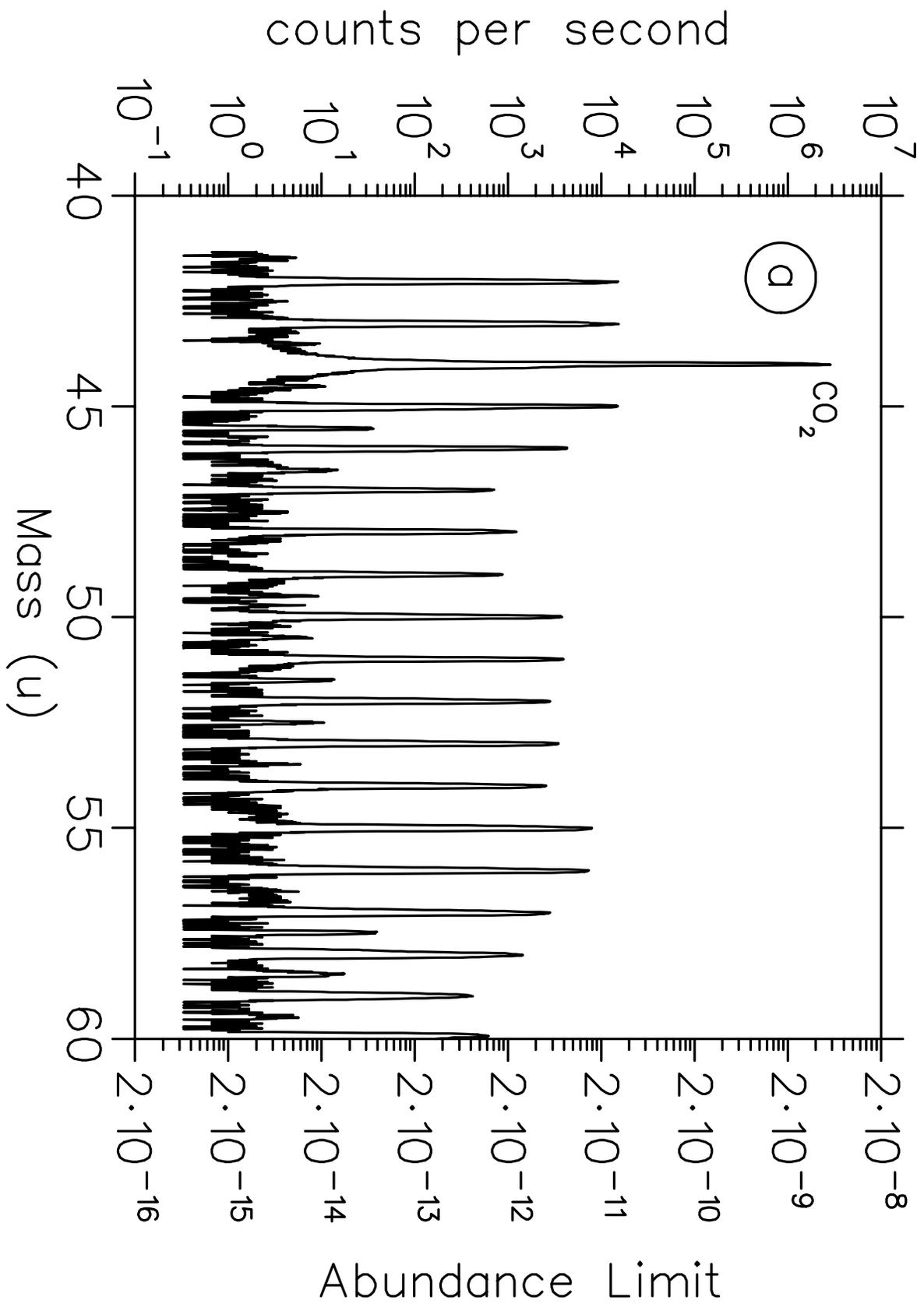
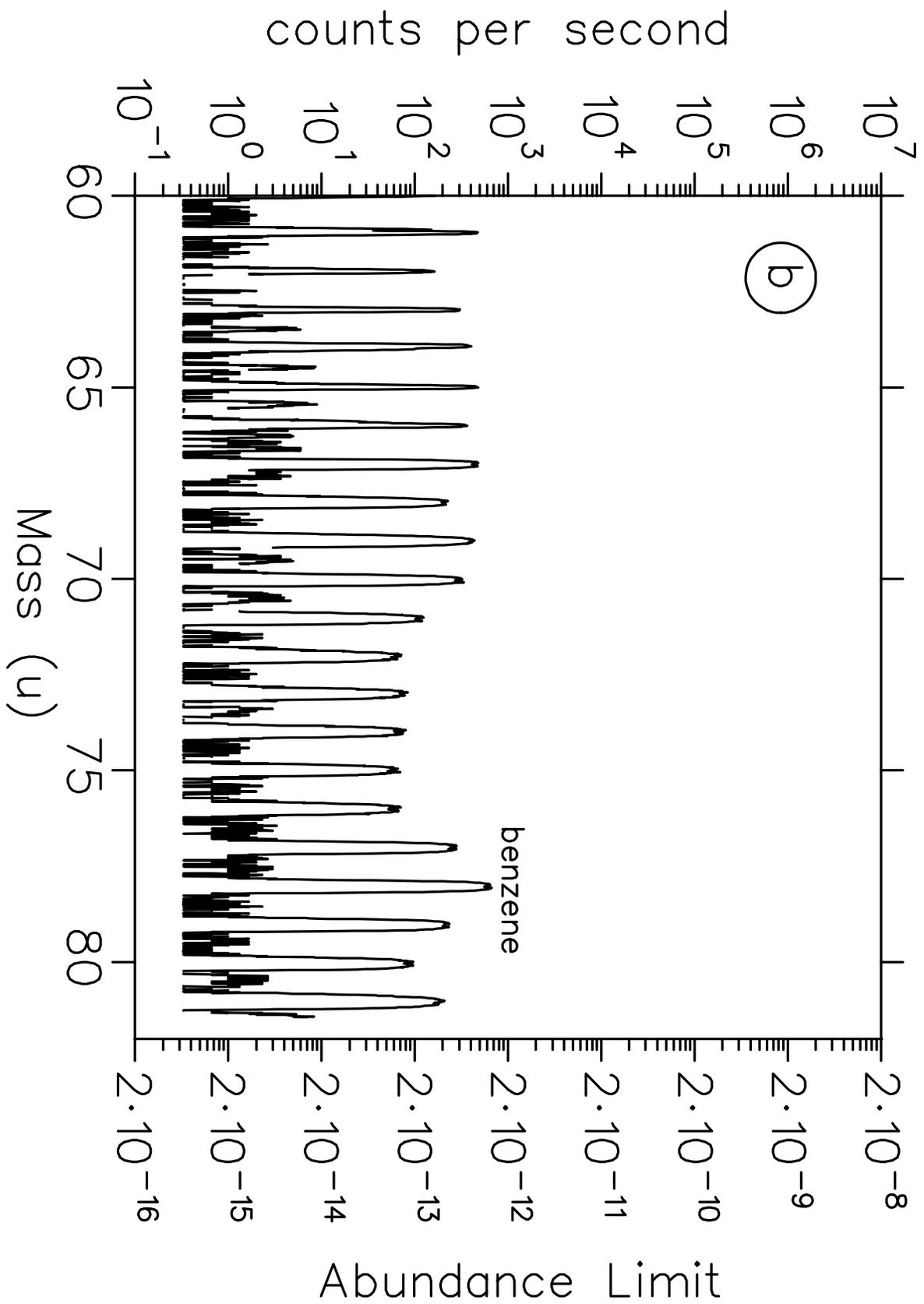
\end{document}